\documentclass[aps,prd,preprintnumbers,amsmath,amssymb,nofootinbib,11pt]{revtex4}
\usepackage{eepic}
\usepackage{indentfirst}
\usepackage{mathrsfs}
\usepackage{fancyhdr}
\usepackage{ulem}
\usepackage{float}
\usepackage{graphicx}
\usepackage[
colorlinks=true, linkcolor=black, breaklinks=true, urlcolor=blue,
citecolor=green]{hyperref}
\usepackage{epstopdf}
\usepackage{bm,bbm}

\usepackage{tabularx}
\usepackage{subfigure}
\usepackage{epsfig}
\usepackage{color}
\usepackage{slashed}
\usepackage{hyperref}

\newcommand{\be}{\begin{equation}}
\newcommand{\ee}{\end{equation}}
\newcommand{\bqa}{\begin{eqnarray}}
\newcommand{\eqa}{\end{eqnarray}}
\renewcommand{\L}{\mathscr{L}}

\newcommand{\bra}{\langle}
\newcommand{\ket}{\rangle}
\newcommand{\nn}{\nonumber}

\newcommand{\GeV}{\,\text{GeV}}
\renewcommand{\vec}[1]{\mathbf{#1}}

\begin{document}
\pagestyle{plain}

\title {\boldmath Chromopolarizabilities of bottomonia from the $\Upsilon(2S,3S,4S) \to \Upsilon(1S,2S)\pi\pi$ transitions  }

\author{ Yun-Hua~Chen$^a$}\email{yhchen@ustb.edu.cn}
\author{ Feng-Kun Guo$^{b,c}$}\email{ fkguo@itp.ac.cn }
\affiliation{${}^a$School of Mathematics and Physics, University
of Science and Technology Beijing, Beijing 100083, China\\
                         ${}^b$CAS Key Laboratory of Theoretical Physics,
             Institute of Theoretical Physics, Chinese Academy of Sciences,
Beijing 100190, China \\
             ${}^c$School of Physical Sciences, University of Chinese Academy of
             Sciences, Beijing 100049, China      }

\begin{abstract}

The dipion transitions $\Upsilon(2S,3S,4S) \to \Upsilon(1S,2S)\pi\pi$ are systematically studied by considering the
mechanisms of the hadronization of soft gluons, exchanging the  bottomoniumlike $Z_b$ states, and the bottom-meson loops. The
strong pion-pion final-state interaction, especially including the
channel coupling to $K\bar{K}$ in the $S$-wave, is taken into
account in a model-independent way using the dispersion theory.
Through fitting to the available experimental data, we extract values of
the transition chromopolarizabilities $|\alpha_{\Upsilon(mS)\Upsilon(nS)}|$, which measure the chromoelectric couplings of the bottomonia with soft gluons.
It is found that the $Z_b$ exchange has a slight impact on the extracted chromopolarizablity values, and the obtained $|\alpha_{\Upsilon(2S)\Upsilon(1S)}|$ considering the $Z_b$ exchange is $(0.29\pm 0.20)~\text{GeV}^{-3}$. Our results could be useful in studying the interactions
of bottomonium with light hadrons.

\end{abstract}

\maketitle

\newpage

\section{Introduction}

The chromopolarizability of a heavy  quarkonium state parametrizes the effective interaction of the quarkonium with soft gluons, and it is
an important quantity in describing the interactions of quarkonium with hadrons~\cite{Voloshin:1980zf,Novikov:1980fa,ms86,sm97,ha99,lr02,gs05,Sibirtsev:2005ex}.
The heavy quarkonium chromopolarizability becomes interesting recently because of two reasons. Firstly, it is relevant for the interpretation of the structures of multiquark hadrons containing a pair of heavy quark and antiquark. In the hadro-quarkonium picture for hidden-flavor tetraquarks and the baryo-quarkonium picture for pentaquarks, the compact heavy quark-antiquark pair is embedded in the light quark matter, and the interaction between these two components takes place via multigluon exchanges.
At reasonable values of the chromopolarizabilities of the charmonia,
several hadro-charmonium bound states and baryo-charmonium bound states are found and identified with certain $XYZ$ states and the $P_c^+$ pentaquark states~\cite{Voloshin:2007dx,Dubynskiy:2008mq,Sibirtsev:2005ex,Eides:2015dtr,Tsushima:2011kh} (a lattice study of the possibility of hadroquarkonium can be found in Ref.~\cite{Alberti:2016dru}).
Also, several hidden-bottom bound states are predicted through the study of the spectrum of the hadro-bottomonium and baryo-bottomonium,
and the emergence of these bound states is sensitive to the value of the bottomonium chromopolarizability~\cite{Ferretti:2018kzy,Anwar:2018bpu}. Secondly, it was suggested that the near-threshold production of heavy quarkonium is sensitive to the trace anomaly contribution to the nucleon mass~\cite{Kharzeev:1995ij}, which may be measured at Jefferson Laboratory and future electron-ion colliders~\cite{Joosten:2018gyo} (for a recent discussion, see Ref.~\cite{Hatta:2018ina}). The suggestion is based on the vector-meson dominance model and the assumption that the nucleon interacts with the heavy quarkonium through the exchange of gluons. We notice that, however, the $\Lambda_c^+D^-$ threhsold is only 116~MeV above the $J/\psi p$ threshold, making the contribution from the $\Lambda_c\bar D$ channel to the $J/\psi p$ near-threshold production nonnegligible. The $\Lambda_b B$ threshold is more than 500~MeV above the $\Upsilon p$ threshold. As a result the $\Upsilon p$ near-threshold photoproduction could be a better process for that purpose, and the  chromopolarizability for the $\Upsilon$ needs to be understood well first.

The diagonal chromopolarizability $\alpha_{QQ}$, with $Q$ representing a heavy quarkonium, cannot be extracted directly from the present experimental data. A possible approach to calculate
$\alpha_{Q Q}$ is based on considering the heavy quarkonia as purely Coulombic systems. This could be a reasonable approximation for the ground state bottomonia, while it is questionable for charmonia and excited bottomonia~\cite{Anwar:2018bpu}. On the other hand, the determination of the nondiagonal (transition) chromopolarizability $\alpha_{Q^\prime Q}\equiv\alpha_{Q^\prime\to Q}$ is of importance since it is natural to expect that each of the diagonal amplitudes should be larger than the nondiagonal amplitude, thus the transition chromopolarizability acts a reference benchmark for either of the diagonal terms~\cite{Voloshin:2004un,Sibirtsev:2005ex}.
Phenomenological value of the bottomonium transition chromopolarizability $\alpha_{\Upsilon(2S)\Upsilon(1S)}$ has been extracted from the process of $\Upsilon(2S) \to \Upsilon(1S) \pi\pi$, and the result is
$|\alpha_{\Upsilon(2S)\Upsilon(1S)}| \approx 0.66$ GeV$^{-3}$~\cite{Voloshin:2004un,Voloshin:2007dx}, where the $\pi\pi$  final-state interaction (FSI) was not considered. Taking account of the $\pi\pi$ $S$-wave FSI in a chiral unitary approach, it is found that the value of $|\alpha_{\Upsilon(2S)\Upsilon(1S)}|$ may be reduced to about $1/3$ of that without the $\pi\pi$ FSI~\cite{Guo:2006ya}. All these previous studies did not consider the effects of the
two bottomoniumlike exotic states $Z_b(10610)$ and $Z_b(10650)$
discovered in channels including $\Upsilon(nS)\pi $
($n=1, 2, 3$) by the Belle Collaboration in 2011~\cite{Belle2011:1,Belle2012:1}. In our previous studies which focus on describing the $\pi\pi$ invariant mass spectrum, we found that
the $Z_b(10610)^\pm$ and $Z_b(10650)^\pm$ bottomonium-like states, though being virtual, play a special role in the hadronic transitions
$\Upsilon(4S,3S,2S) \to \Upsilon(nS) \pi \pi $~\cite{Chen2016,Chen:2016mjn}. Thus the discovery of two $Z_b$ resonances necessitates a reanalysis of the transition chromopolarizabilities
in the dipion transitions between the $\Upsilon$ states. In addition, there have been new measurements after our analysis in Refs.~\cite{Chen2016,Chen:2016mjn} by the Belle Collaboration with statistics higher than before, and especially they measured the angular distributions
of the $\Upsilon(4S) \to \Upsilon(1S,2S) \pi \pi $ transitions for the first time~\cite{Belle2017}. These new data help us to perform a comprehensive analysis of the $\Upsilon(4S,3S,2S) \to \Upsilon(nS) \pi \pi $ processes.

Since the $\Upsilon(4S)$ meson is above the $B\bar{B}$ threshold and
decays predominantly to $B\bar{B}$, the intermediate bottom-meson loops
need to be taken into account in the analysis of the $\Upsilon(4S,3S,2S) \to \Upsilon(nS) \pi \pi $ processes. The $\pi\pi$ FSI plays an important role in the heavy quarkonium transitions and modifies the value of transition chromopolarizability significantly~\cite{Guo:2006ya,Chen:2019hmz}, and it is thus necessary to account for its effects properly. In this work we will use the dispersion theory in the form of modified Omn\`es
solutions to consider the FSI.\footnote{The $\pi\pi$ FSI may also be implemented through the generalized distribution amplitude as discussed in Refs.~\cite{Diehl:1998dk,Diehl:2000uv}.} The sum of
the $Z_b$-exchange mechanism and the bottom meson loops provide the left-hand-cut contribution to
the dispersion integral representation~\cite{Chen2016,Chen:2016mjn}.

This paper is organized as follows. In Sec.~\ref{theor}, we
introduce the theoretical framework. In Sec.~\ref{pheno}, we present the fit
results and discuss the phenomenology.
Summary and conclusions are given in Sec.~\ref{conclu}.

\section{Theoretical framework}
\label{theor}

First we define the the Mandelstam variables for the decay process $\Upsilon(mS)(p_a) \to \Upsilon(nS)(p_b) \pi(p_c)\pi(p_d)$
\begin{align}
s &= (p_c+p_d)^2 , \qquad
t=(p_a-p_c)^2\,, \qquad u=(p_a-p_d)^2\,,
\end{align}
where $p_{a,b,c,d}$ are the corresponding four-momenta.

The standard mechanism for these transitions was
thought to be the emission of soft gluons from compact bottomonium, followed by their hadronization
into two pions. For the bottomnium size being much smaller than the gluon wave length, such a mechanism may be calculated by the nonperturbative quantum chromodynamics (QCD) multipole expansion method, and the amplitude for the dipion transition between $S$-wave states $A$ and $B$ of heavy quarkonium
can be written as~\cite{Novikov:1980fa,Voloshin:1982ij}
\begin{eqnarray}\label{eq.MultipoleAmplitude}
M_{AB}&=& 2\sqrt{m_Am_B}\alpha_{AB}\left\langle
\pi^+(p_c)\pi^-(p_d)\left|\frac{1}{2}\vec{E}^a\cdot \vec{E}^a\right|0\right\rangle \nonumber \\
&=&\frac{8\pi^2}{b}\sqrt{m_A m_B}\alpha_{AB}(\kappa_1 p^0_c p^0_d-\kappa_2 p^i_c p^i_d),
\end{eqnarray}
where the factor $2\sqrt{m_Am_B}$ appears due to the relativistic
normalization of the decay amplitude $M_{AB}$, $\alpha_{AB}$ is the transition chromopolarizability, $\vec{E}^a$ denotes
the chromoelectric field, and the second line is from trace anomaly. Here, $b=\frac{11}{3}N_c-\frac{2}{3}N_f$ refers to the first coefficient of the QCD beta function,
with $N_c=3$ and $N_f=3$ the numbers of colors and of light flavors, respectively, and $\kappa_1$ and $\kappa_2$ are not independent as $\kappa_1=2-9\kappa/2$ and $\kappa_2=2+3\kappa/2$, where the parameter $\kappa$ can be determined from fitting to data.
The above expression can be reproduced by constructing a chiral
effective Lagrangian for the contact $\Upsilon(mS) \to \Upsilon(nS) \pi\pi$
transition. Since the spin-dependent interactions are suppressed for heavy quarks, the heavy quarkonia
can be expressed in term of spin multiplets, and one has $J \equiv \vec{\Upsilon}
\cdot \boldsymbol{\sigma}+\eta_b$, where $\boldsymbol{\sigma}$
contains the Pauli matrices and $\vec\Upsilon$ and $\eta_b$
annihilate the $\Upsilon$ and $\eta_b$ states, respectively (see, e.g., Ref.~\cite{Guo2011}). The effective
Lagrangian, at the leading order in the chiral as well as the heavy-quark
nonrelativistic expansion, reads~\cite{Mannel,Chen2016,Chen:2016mjn}
\begin{equation}\label{LagrangianUpUppipi}
\L_{\Upsilon\Upsilon^{\prime}\Phi\Phi} = \frac{c_1}{2}\bra J^\dagger
J^\prime \ket \bra u_\mu u^\mu\ket +\frac{c_2}{2}\bra J^\dagger
J^\prime \ket \bra u_\mu u_\nu\ket v^\mu v^\nu +\mathrm{h.c.} \,,
\end{equation}
where $u_\mu=-\partial_\mu \Phi/F_\pi+\mathcal{O}(\Phi^3)$, with $\Phi=\boldsymbol{\tau}\cdot\boldsymbol{\pi}$ the pion fields, $\boldsymbol{\tau}$ the Pauli martices, and $F_\pi=92.1$~MeV the pion decay constant, is the axial current collecting
the Goldstone bosons (pions) of the spontaneous breaking of chiral symmetry, and
$v^\mu=(1,\vec{0})$ is the velocity of the heavy quark.
The contact term amplitude obtained by using the chiral  Lagrangian
in Eq.~\eqref{LagrangianUpUppipi} reads
\begin{equation}
\label{eq.ChiralAmplitude}
M(s,t,u)
= -\frac{4}{F_\pi^2}\left( c_1 p_c\cdot p_d +c_2 p_c^0 p_d^0\right)\,.
\end{equation}
Matching the amplitude in Eq.~\eqref{eq.MultipoleAmplitude} to that in Eq.~\eqref{eq.ChiralAmplitude}, we
can express the chiral low-energy coupling constants in terms of the chromopolarizability $\alpha_{AB}$ and the parameter $\kappa$,
\begin{align}
c_1&= -\pi^2\sqrt{m_{\Upsilon^\prime}m_\Upsilon}F_\pi^2\alpha_{\Upsilon^\prime\Upsilon}\frac{4+3\kappa}{b} , \nn\\
c_2&= 12\pi^2\sqrt{m_{\Upsilon^\prime}m_\Upsilon}F_\pi^2\alpha_{\Upsilon^\prime\Upsilon}\frac{\kappa}{b}\,.\label{eq.Matching}
\end{align}

\begin{figure}[tb]
\centering
\includegraphics[width=\linewidth]{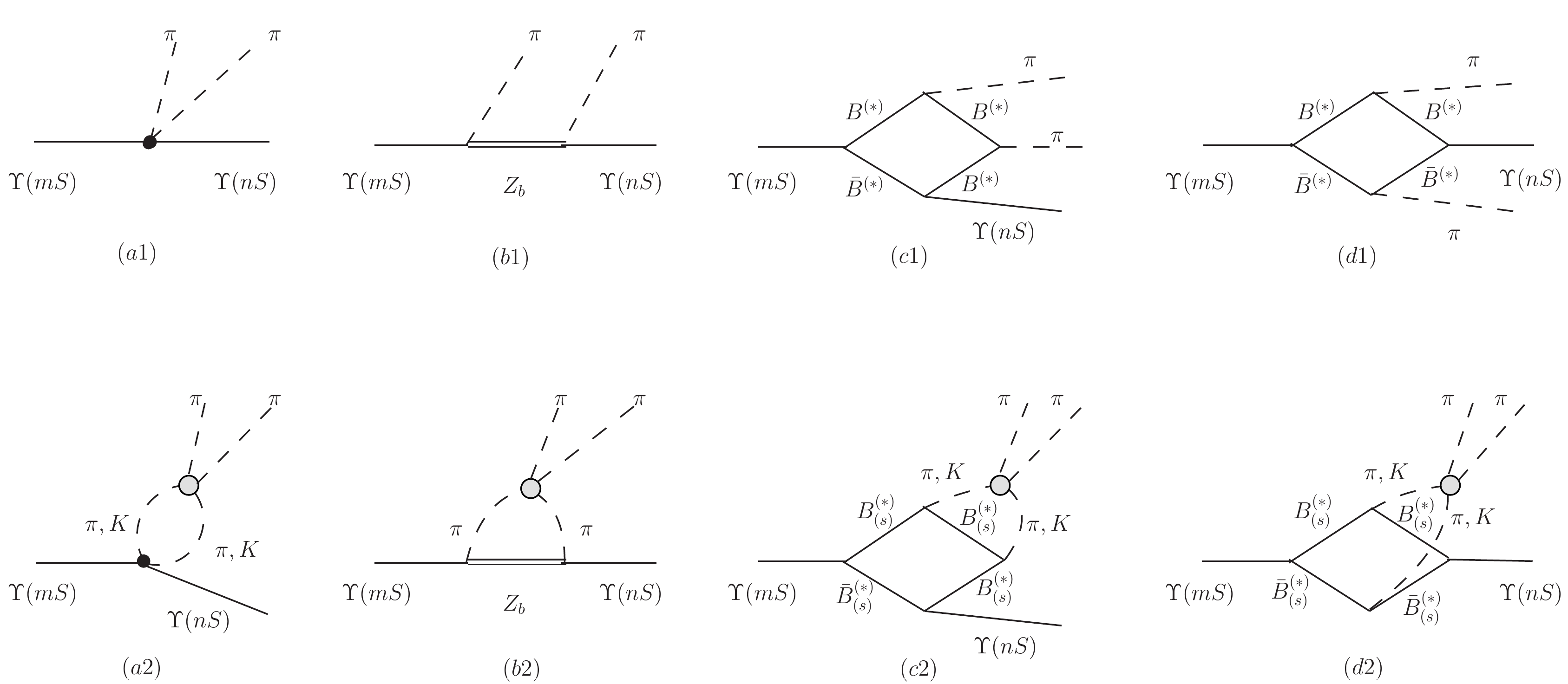}
\caption{Feynman diagrams considered for the $\Upsilon(mS)
\rightarrow \Upsilon(nS) \pi \pi $ processes. The crossed diagrams
of (b1), (c1), (b2), and (c2) are not shown explicitly. The gray
blob denotes the FSI.
}\label{fig.FeynmanDiagram}
\end{figure}

In addition to the multipole contribution $\Upsilon(mS) \to \Upsilon(nS)+\text{gluons} \to \Upsilon(nS)\pi\pi$ which has been parametrized into the chiral contact terms in Eq.~\eqref{LagrangianUpUppipi}, we also take into account the mechanisms of the $Z_b$-exchange and the bottom meson loops. In addition, for a complete theoretical treatment of the dipion transitions, as mentioned above, the $\pi\pi$ FSI needs to be taken into account as well. It is considered using the dispersion theory which has been fully described in our previous papers~\cite{Chen:2016mjn,Chen2016} (the left-hand cuts from the bottom-meson loops are not considered in Ref.\cite{Chen2016}),
and we only list the relevant Lagrangians for  defining the parameters in the following. The relevant Feynman diagrams for the $\Upsilon(mS)
\rightarrow \Upsilon(nS) \pi \pi $ processes are displayed in Fig.~\ref{fig.FeynmanDiagram}.

The leading order chiral Lagrangian for the  $Z_b \Upsilon\pi$ interaction reads~\cite{Guo2011}
\be\label{LagrangianZbUppi} \L_{Z_b\Upsilon\pi} =
\sum_{j=1,2}\sum_{n}C_{Z_{bj}\Upsilon(lS)\pi} \Upsilon^i(nS) \bra
{Z^i_{bj}}^\dagger u_\mu \ket v^\mu +\mathrm{h.c.} \,,
\ee
where
$Z_{b1}$ and $Z_{b2}$ are used to refer to the $Z_b(10610)$ and
$Z_b(10650)$, respectively. The mass difference between the two $Z_b$ states is
much smaller than the difference between their masses and the $\Upsilon(nS)\pi$ thresholds; they have
the same quantum numbers and thus the same coupling structure as dictated by Eq.~\eqref{LagrangianZbUppi}.
As a result, they can hardly be distinguished from each other in the processes studied here, so we only
use one effective $Z_b$ state, the $Z_b(10610)$, to include the $Z_b$ effects as done in Refs.~\cite{Chen:2016mjn,Chen2016}.

To calculate the box diagrams, we need the effective Lagrangian for the
coupling of the bottomonium fields to the bottom and
antibottom mesons~\cite{Guo2009:PRL},
\begin{equation}\label{LagrangianJHH}
\L_{JHH}=\frac{i\, g_{JHH}}{2}\bra J^\dag H_a
\boldsymbol{\sigma}\cdot \!\overleftrightarrow{\partial}\!
\bar{H}_a\ket + {\rm h.c.}\,,
\end{equation}
and the coupling of the Goldstone
bosons to the bottom and antibottom mesons~\cite{Burdman:1992gh,Wise:1992hn,Yan:1992gz,Casalbuoni:1996pg,Mehen2008}
\begin{equation}\label{LagrangianHHPhi}
\L_{HH\Phi}= \frac{g_\pi}{2} \bra \bar{H}_a^\dagger  \boldsymbol{\sigma} \cdot \vec{u}_{ab}
\bar{H}_b\ket -\frac{g_\pi}{2} \bra H_a^\dagger H_b
\boldsymbol{\sigma} \cdot \vec{u}_{ba} \ket,
\end{equation}
where $H_a=\vec{V}_a \cdot \boldsymbol{\sigma}+P_a$ with $\boldsymbol{\sigma}$ the Pauli matrices and
$P_a(V_a)=(B^{(*)-},\bar{B}^{(*)0},\bar{B}_s^{(*)0})$~\cite{Mehen2008}.
We use $g_\pi=0.5$ for the axial coupling from a recent lattice QCD
calculation~\cite{Bernardoni:2014kla}.

\section{Phenomenological discussion}\label{pheno}

For each $\Upsilon(mS) \to \Upsilon(nS) \pi\pi$ transition, the unknown parameters include the chromopolarizability $\alpha_{\Upsilon(mS)\Upsilon(nS)}$, the parameter $\kappa_{\Upsilon(mS)\Upsilon(nS)}$, the product of couplings for the effective $Z_b$-exchange $C_{Z_b\Upsilon(mS)\pi}C_{Z_b\Upsilon(nS)\pi}$, and the product of couplings for the box diagrams $g_{JHH(mS)}g_{JHH(nS)}$.
The value of
$g_{JHH(4S)}$ can be extracted from the measured open-bottom decay widths of the
$\Upsilon(4S)$, $g_{JHH(4S)}=1.43\GeV^{-3/2}$. The unknown couplings $g_{JHH(1S)}$, $g_{JHH(2S)}$ and $g_{JHH(3S)}$ will be fixed from  simultaneously fitting to the experimental data of the $\pi\pi$ invariant mass distributions and the helicity angular distributions of the
$\Upsilon(2S) \to \Upsilon(1S) \pi\pi$, $\Upsilon(3S) \to \Upsilon(1S) \pi\pi$, and $\Upsilon(4S) \to \Upsilon(1S,2S) \pi\pi$ processes.

\begin{figure}[tbhp]
\centering
\includegraphics[height=18cm,width=16cm]{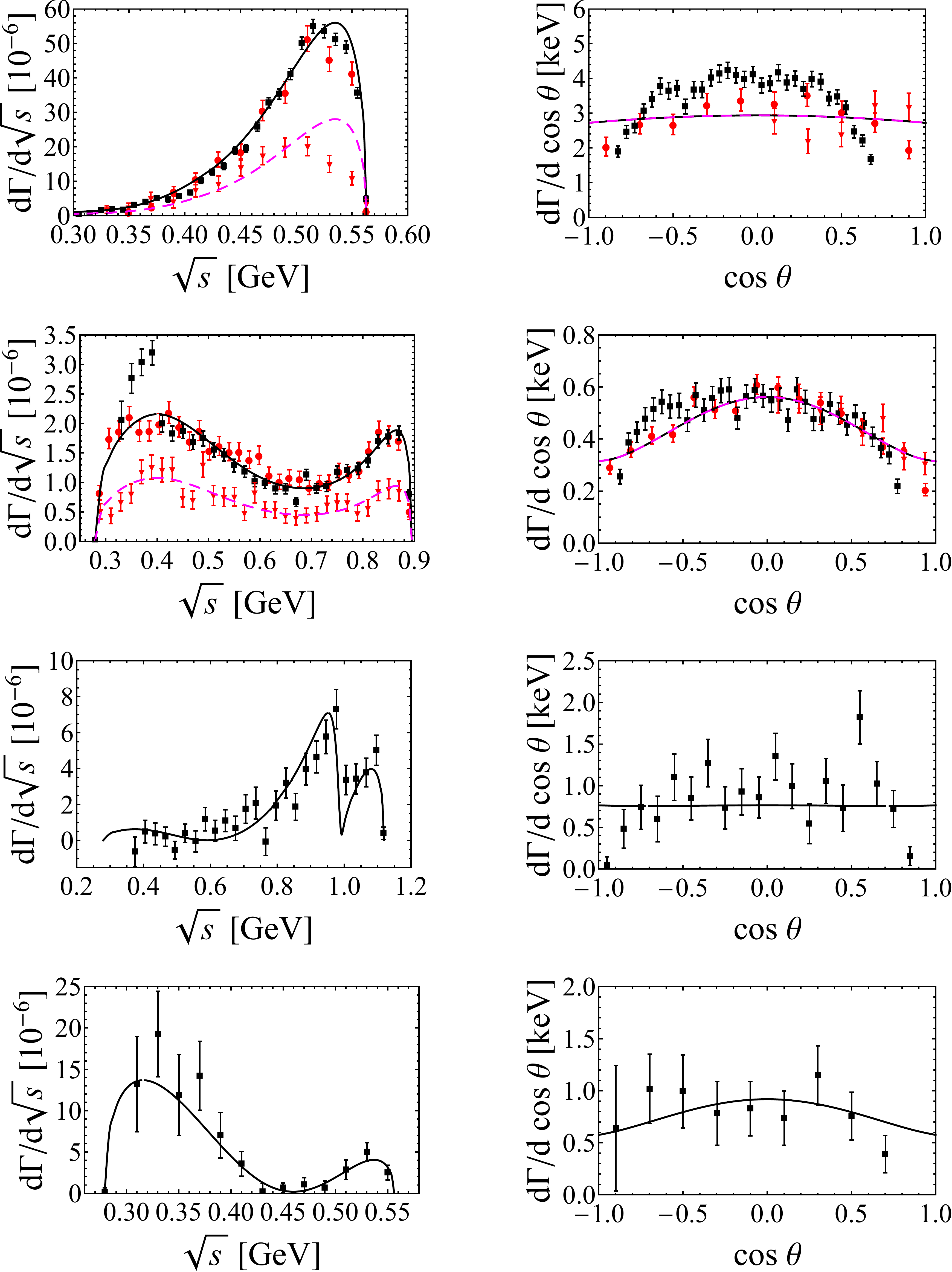}
\caption{Fit results for the decays $\Upsilon(2S) \to \Upsilon(1S)
\pi\pi $, $\Upsilon(3S) \to \Upsilon(1S) \pi\pi $, $\Upsilon(4S) \to \Upsilon(1S) \pi^+\pi^- $, and $\Upsilon(4S)
\to \Upsilon(2S) \pi^+\pi^- $ (from top to bottom). The left panels
display the $\pi\pi$ invariant mass spectra, while the right panels
show the $\cos\theta$ distributions. The solid squares denote the
charged decay mode data from the Belle Collaboration~\cite{Belle2017}. The solid circles and solid triangles
denote the charged and neutral decay mode data, respectively, from the CLEO Collaboration~\cite{CLEO2007}. The solid black
and solid magenta lines show the best fit results for charged- and
neutral-pion final states.
}\label{fig.mSnSFitResults}
\end{figure}

\begin{figure}[h]
\centering
\includegraphics[height=18cm,width=16cm]{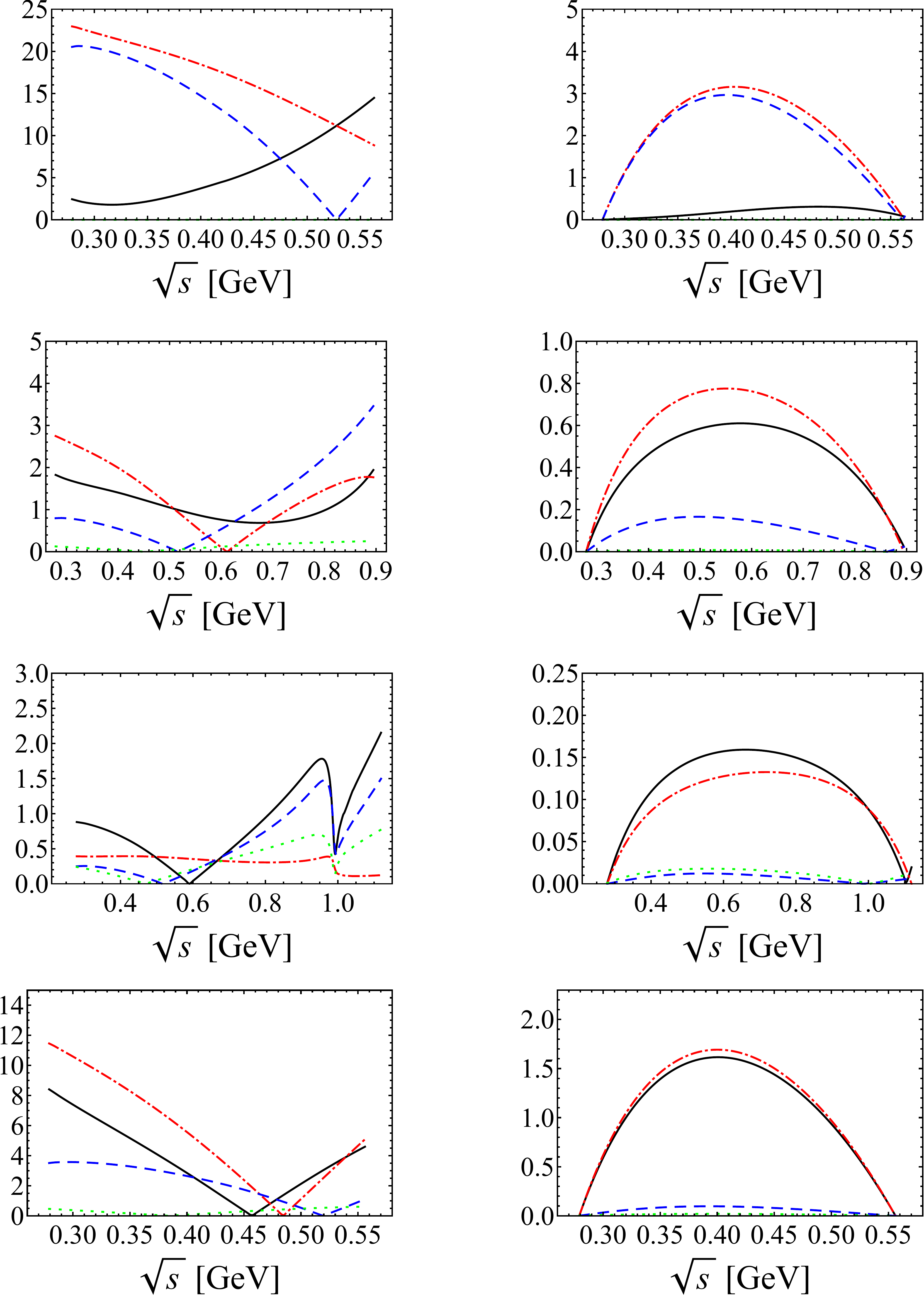}
\caption{Moduli of the $S$- (left) and $D$-wave (right) amplitudes
in the decays $\Upsilon(2S) \to \Upsilon(1S)
\pi\pi $, $\Upsilon(3S) \to \Upsilon(1S) \pi\pi $, $\Upsilon(4S) \to \Upsilon(1S) \pi^+\pi^- $, and $\Upsilon(4S)
\to \Upsilon(2S) \pi^+\pi^- $ (from top to bottom). The black solid
lines represent our best fit results, while the red dot-dashed, blue
dashed, and green dotted lines correspond to the contributions from
the chiral contact terms, the $Z_b$, and the box diagrams,
respectively. }\label{fig.SwaveDwaveAmplitudes}
\end{figure}

The results of the best fit are shown as the solid black
(solid magenta) curves for the $\pi^+\pi^-$ ($\pi^0\pi^0$) mode in Figs.~\ref{fig.mSnSFitResults}.
The fitted parameters as well
as the $\chi^2/(\text{number of events})$ for each $\Upsilon(mS) \to \Upsilon(nS) \pi\pi$ transition are given in Table~\ref{tablepar1}.
Using the central values of the parameters in the best fit, in
Fig.~\ref{fig.SwaveDwaveAmplitudes} we plot the moduli of the $S$-
and $D$-wave amplitudes from the chiral contact terms, the effective $Z_b$-exchange, and the box graphs for each $\Upsilon(mS) \to \Upsilon(nS) \pi\pi$ transition.

\begin{table}[bth]
\caption{\label{tablepar1} Fit parameters from the best simultaneous fit of the
$\Upsilon(mS) \to \Upsilon(nS) \pi\pi$ $(n<m\leq 4)$ processes.}
\renewcommand{\arraystretch}{1.1}
\begin{ruledtabular}
\begin{tabular}{l|cccc}
         & $~\Upsilon(2S) \to \Upsilon(1S) \pi\pi ~$
         & $~\Upsilon(3S) \to \Upsilon(1S) \pi\pi ~$
         & $~\Upsilon(4S) \to \Upsilon(1S) \pi^+\pi^- ~$
         & $~\Upsilon(4S) \to \Upsilon(2S) \pi^+\pi^- ~$
         \\
\hline
$|\alpha_{\Upsilon(mS)\Upsilon(nS)}|~[\text{GeV}^{-3}]$  & $0.29\pm 0.20$& $0.06\pm 0.03$ &   $ (5.4\pm 3.5)\times 10^{-4}$           & $0.43\pm 0.01$\\
$\kappa_{\Upsilon(mS)\Upsilon(nS)}$   & $ 1.52\pm 1.17$& $ 0.34\pm 0.19$&   $ -3.3\pm 2.1$  & $0.53\pm 0.02$  \\
\hline
 ${\chi^2}/(\text{number of events})$ &  ${794.7}/{98}$&  ${288.4}/{151}$&  ${75.3}/{43}$   &  ${14.7}/{23}$  \\
\botrule
 & $ |C_{Z_{b1}\Upsilon(1S)\pi}|$& $ |C_{Z_{b1}\Upsilon(2S)\pi}|$& $ |C_{Z_{b1}\Upsilon(3S)\pi}|$ &$ |C_{Z_{b1}\Upsilon(4S)\pi}|$ \\
 \hline
 & $(5.7\pm 0.2)\times 10^{-2}$& $1.6\pm 0.1$& $ (2.1\pm 0.1)\times 10^{-2}$ &$ (3.3\pm 0.1)\times 10^{-3}$ \\
\botrule
  & $|g_{JHH(1S)}|~[\text{GeV}^{-{3}/{2}}]$& $|g_{JHH(2S)}|~[\text{GeV}^{-{3}/{2}}]$& $|g_{JHH(3S)}|~[\text{GeV}^{-{3}/{2}}]$ &\\
 \hline
  & $(4.1\pm 0.2)\times 10^{-5}$ & $(2.7\pm 0.8)\times 10^{-4}$ & $1.4\pm 5.1$ &\\
\end{tabular}
\end{ruledtabular}
\renewcommand{\arraystretch}{1.0}
\end{table}

Several remarks about the fitting results are in order:
\begin{enumerate}

\item
For the $\Upsilon(2S)\to\Upsilon(1S)\pi\pi$ process, there are large discrepancies between our theoretical output and the angular distribution data measured by Belle.
As shown in Fig.~\ref{fig.SwaveDwaveAmplitudes}, for the dominant chiral contact terms and the $Z_b$-exchange term,
their $D$-wave components are about one order of magnitude smaller than the corresponding $S$-wave ones. Thus, a rather flat angular distribution is expected in our scheme, which agrees with the CLEO measurement, but not with the Belle measurement. In addition, one notices that in the $\psi^\prime \to J/\psi\pi\pi$ transition,
a rather flat angular distribution was observed experimentally~\cite{Ablikim:2006bz}.

For the transition chromopolarizability, considering only the multipole contribution $\Upsilon(mS) \to \text{gluons} + \Upsilon(nS) \to \Upsilon(nS)\pi\pi$ (i.e., the chiral contact terms), the value without FSI was obtained as
$|\alpha_{\Upsilon(2S)\Upsilon(1S)}|\approx 0.66~ \text{GeV}^{-3}$~\cite{Voloshin:2004un,Voloshin:2007dx}, and the value including the $\pi\pi$ FSI in a chiral unitary approach is $|\alpha_{\Upsilon(2S)\Upsilon(1S)}|=0.24\pm 0.01~ \text{GeV}^{-3}$~\cite{Guo:2006ya}.
As shown in Table~\ref{tablepar1}, the effects of $Z_b$-exchange and the box diagrams modify the value of the chromopolarizability slightly, and now it is $|\alpha_{\Upsilon(2S)\Upsilon(1S)}|=0.29\pm 0.20~ \text{GeV}^{-3}$, which agrees with the result in Ref.~\cite{Guo:2006ya} within errors.

For the parameter $\kappa$, one observes that the value from our fit $\kappa_{\Upsilon(2S)\Upsilon(1S)}=1.52\pm 1.17$, carrying a sizeable uncertainty. Its central value is larger than the result $\kappa_{\Upsilon(2S)\Upsilon(1S)}=0.342^{+0.015}_{-0.017}$ in Ref.~\cite{Pineda:2019mhw} using QCD multipole expansion, which was obtained from fitting to the $\pi\pi$ differential decay width spectrum of $\Upsilon(2S)\to\Upsilon(1S)\pi\pi$ using a chiral effective Lagrangian as in Ref.~\cite{Mannel:1995jt}. There are four differences between our treatment and that in Ref.~\cite{Pineda:2019mhw}: (1) we have considered $\pi\pi$ FSI, (2) we have considered the $Z_b$, (3) we have considered the bottom-meson box diagrams, and (4) we dropped the term proportional to the quark mass matrix in the chiral Lagrangian since the same term will introduce a $\Upsilon(2S)\Upsilon(1S)$ mixing by virtual of chiral symmetry and should be eliminated upon diagonalizing the mass matrix for the $\Upsilon$ states as argued in~\cite{Chen2016}. Among them, (2) and (3) are non-multipole effects, and (1) is mandatory in particular for the $\pi\pi$ $S$ wave since the $f_0(500)$ resonance is located in this energy range. Our earlier analysis in Ref.~\cite{Chen2016}, where the bottom-meson box diagrams were not considered, led to a value of $-0.13 \pm 0.25$ for $\kappa_{\Upsilon(2S)\Upsilon(1S)}$.

One observes the following hierarchy from our fit:
$|\alpha_{\Upsilon(4S)\Upsilon(1S)}| \ll |\alpha_{\Upsilon(3S)\Upsilon(1S)}| \ll |\alpha_{\Upsilon(2S)\Upsilon(1S)}|
\lesssim |\alpha_{\Upsilon(4S)\Upsilon(2S)}|$, which agrees with the expectation in Ref.~\cite{Voloshin:2004un}. This may be qualitatively understood from the node
structure of the $\Upsilon(nS)$ wave functions~\cite{TMYan1980,Kuang1981}: for the processes with
the same final $\Upsilon$ state, the larger the difference between
the principal quantum numbers, the smaller the gluonic matrix
elements and thus the magnitude of the transition chromopolarizabilities.

\item
For the $\Upsilon(3S) \to \Upsilon(1S) \pi\pi$ process, one observes that the two-hump structure of the
$\pi\pi$ mass spectrum and the angular distribution can be well reproduced. One notices that there is a jump at around 0.35~GeV in the Belle data, which, however, is dubious since there is no threshold or any other singularity in that region. The Belle data points below 0.35~GeV contribute sizeably to the value of $\chi^2$.

\item
For the $\Upsilon(4S) \to \Upsilon(1S) \pi^+\pi^-$ process,
the dipion mass spectrum indeed has a dip around 1 GeV in the new Belle data, which has been
predicted due to the presence of the $f_0(980)$~\cite{Chen:2016mjn}. We further notice that now the
data points left to the $f_0(980)$ are the highest ones and the line shape there is lifted up mainly by the $Z_b$-exchange mechanism. This feature can be seen in Fig.~\ref{fig.SwaveDwaveAmplitudes},
where one observes that for the dominant $S$-wave amplitudes, the $Z_b$ exchange plays a major role in the energy range around 0.95 GeV. Thus, the effective couplings of $Z_b$ to $\Upsilon(4S)\pi$ and $\Upsilon(1S)\pi$ are better constrained compared with our previous study~\cite{Chen:2016mjn}. For the angular distribution, the theoretical prediction is very flat since the $D$-wave contribution is much smaller than the $S$-wave one.

\item
For the $\pi\pi$ mass spectrum of the $\Upsilon(4S) \to \Upsilon(2S) \pi^+\pi^-$ process, the new Belle data show a two-peak structure as
in the old BABAR data~\cite{BABAR2006}, while a distinct difference is that in the Belle data the dip approaches zero inside the physical region. Since the
chiral contact amplitude contains a zero in this energy range, the $\pi\pi$ mass spectrum of the Belle data can be described well even by only
including the chiral contact terms with FSI as we have checked. As a result, the value of $|g_{JHH(2S)}|$ turns out to be smaller than that determined in Ref.~\cite{Chen:2016mjn} where the BaBar data with larger uncertainties~\cite{BABAR2006} were used. In the BaBar data, the dip at around 0.45~GeV is higher, leading to a larger value of $|g_{JHH(2S)}|$.

\item

The branching fractions of the
decays of both $Z_b$ states into $\Upsilon(nS)\pi$
$(n\leq3)$ have been reported by Belle in Ref.~\cite{Garmash:2015rfd}, where the $Z_b$ line shapes were fitted using
Breit--Wigner forms. If we naively calculated the partial widths by multiplying these branching fractions by the measured widths of the two $Z_b$ states,
we would obtain the $Z_{bi}\Upsilon(nS)\pi$ coupling strengths\footnote{In~\cite{Chen2016}, the nonrelativistic normalization factor of $\sqrt{M}$ for heavy mesons has been absorbed into the coupling constants, so the coupling constants therein differ from the corresponding ones in Eq.~\eqref{eq.CZvalue} by a factor of $\sqrt{M_{Z_{bi}}M_{\Upsilon(mS)}}$.}
\begin{align}\label{eq.CZvalue}
|C_{Z_{b1} \Upsilon(1S)\pi}^\text{naive}|&=(3.1 \pm 0.5)\times 10^{-3}, \nonumber\\
|C_{Z_{b2} \Upsilon(1S)\pi}^\text{naive}|&=(1.3 \pm 0.3)\times 10^{-3}, \nonumber\\
|C_{Z_{b1} \Upsilon(2S)\pi}^\text{naive}|&=(2.1 \pm 0.3)\times 10^{-2}, \nonumber\\
|C_{Z_{b2} \Upsilon(2S)\pi}^\text{naive}|&=(0.9 \pm 0.2)\times 10^{-2}, \nonumber\\
|C_{Z_{b1} \Upsilon(3S)\pi}^\text{naive}|&=(5.8 \pm 0.9)\times 10^{-2}, \nonumber\\
|C_{Z_{b2} \Upsilon(3S)\pi}^\text{naive}|&=(3.0 \pm 0.5)\times 10^{-2}\,,
\end{align}
by using
\begin{equation}\label{eq.CZ}
|C_Z|=\Bigg\{\frac{4\pi F_\pi^2 m_{Z_b} \Gamma_{Z_b \to
\Upsilon\pi}}{m_\Upsilon|\vec{p}_f|\big(m_\pi^2+\vec{p}_f ^2\big)}
\Bigg\}^{\frac{1}{2}},
\end{equation}
where $|\vec{p}_f|\equiv
\lambda^{1/2}\big(m_{Z_b}^2,m_\Upsilon^2,m_\pi^2\big)/(2m_{Z_b})$.
One observes that our results of the coupling strengths for $|C_{Z_{b1}\Upsilon(1S)\pi}|$
and $|C_{Z_{b1}\Upsilon(2S)\pi}|$ in Table~\ref{tablepar1}
are about
one or two orders of magnitude larger than those listed above, and the values of $|C_{Z_{b1}\Upsilon(3S)\pi}|$ in Table~\ref{tablepar1} and
in Eq.~\eqref{eq.CZvalue} are of the same order of magnitude. Notice that as analyzed in our previous work~\cite{Chen2016}, the Breit--Wigner parameterization used Ref.~\cite{Garmash:2015rfd} is not the appropriate
way for describing the $Z_b$ line shapes; the $Z_b$
states are very close to the $B^{(*)}\bar B^*$ thresholds, and thus a
Flatt\'e parameterization should be used, which would lead to much larger partial
widths into $\Upsilon(nS)\pi$, and thus the relevant coupling strengths. For more details, we refer to Ref.~\cite{Chen2016}.
In addition, since both $Z_b$ states are well above the $\Upsilon(4S)$ mass, and their effects in the dipion transitions can be hardly distinguished from each other~\cite{Chen2016}, thus we have included only one effective $Z_b$ state in our framework. The so-obtained
coupling strengths $|C_{Z_{b1}\Upsilon(lS)\pi}|$ in Table~\ref{tablepar1} should be understood as effectively containing effects from both of the $Z_b(10610)$
and $Z_b(10650)$ states. Nevertheless, even taking the above two facts into account, the value of
$|C_{Z_{b1}\Upsilon(2S)\pi}|$ in Table~\ref{tablepar1} is too large since it would lead to a partial width of the GeV order using Eq.~\eqref{eq.CZvalue}. Notice that the Belle data of the $\Upsilon(2S) \to \Upsilon(1S) \pi\pi$ process played a crucial role in fixing the value of $|C_{Z_{b1}\Upsilon(2S)\pi}|$, and as mentioned in the first two remarks, the present Belle data on the $\Upsilon(2S,3S) \to \Upsilon(1S)\pi\pi$ transitions have some dubious properties. We except that the future better data of these processes and a proper extraction of the the branching fractions of the
$Z_{bi} \to \Upsilon(nS)\pi$ $(n\leq3)$ decays may help to solve this discrepancy.

\end{enumerate}

\section{Conclusions}
\label{conclu}

We have systemically studied the dipion transitions $\Upsilon(mS) \to \Upsilon(nS) \pi\pi$ with $n<m \leq 4$.
In addition to the multipole contribution $\Upsilon(mS) \to \Upsilon(nS) +\text{gluons} \to \Upsilon(nS) \pi\pi$, the $Z_b$ exchange and bottom-meson loops
are taken into account.
The strong coupled-channel ($\pi\pi$ and $K\bar K$) FSI is considered model-independently by using the dispersion theory.
Through fitting the updated data of the $\pi\pi$ invariant mass spectra and the helicity angular distributions, the values of the transition
the chromopolarizabilities $|\alpha_{\Upsilon(mS)\Upsilon(nS)}|$ are determined. In particular, we find that after including the $Z_b$ exchange and bottom-meson loops the value of $|\alpha_{\Upsilon(2S)\Upsilon(1S)}|$ is determined to be $(0.29\pm 0.20)~\text{GeV}^{-3}$. It is expected in Refs.~\cite{Voloshin:2004un,Sibirtsev:2005ex} that the off-diagonal chromopolarizability should be somewhat smaller than the diagonal one. Within uncertainties, the value of $|\alpha_{\Upsilon(2S)\Upsilon(1S)}|$ from our determination is similar to the diagonal chromopolarizability $|\alpha_{\Upsilon(1S)\Upsilon(1S)}|$, calculated to be in the range of $[0.33,0.47]$~GeV$^{-3}$ in Ref.~\cite{Anwar:2018bpu} and $0.50^{+0.42}_{-0.38}$~GeV$^{-3}$ in Ref.~\cite{Brambilla:2015rqa}, and yet the central value is indeed smaller. The results obtained in this work would be valuable
to understand the chromopolarizabilities of bottomonia, and will have applications for the studies
of light-hadron--bottomonia interactions.

\medskip

\section*{Acknowledgments}

We are grateful to Christoph Hanhart and Bastian Kubis for helpful discussions.
This research is supported in part by the Fundamental Research Funds
for the Central Universities under Grant
No.~FRF-BR-19-001A, by the National Natural Science Foundation of China (NSFC) and  the Deutsche Forschungsgemeinschaft (DFG) through the funds provided to the Sino-German Collaborative Research Center ``Symmetries and the Emergence of Structure in QCD"  (NSFC Grant No. 11621131001, DFG Grant No. TRR110), by the NSFC under Grants No. 11847612 and 11835015, by the Chinese Academy of Sciences (CAS) under Grants No. QYZDB-SSW-SYS013 and XDPB09, and by
the CAS Center for Excellence in Particle Physics (CCEPP).


\end{document}